\documentclass[12pt,thmsa]{article}
\usepackage{amssymb}

\input tcilatex
\QQQ{Language}{American English}

\begin{document}

\begin{center}
\emph{The Self Creation challenge to the cosmological 'concordance model'}

Garth A Barber
\end{center}
\begin{center}
\textit{Kingswood Vicarage, Woodland Way, Tadworth, SURREY KT20 6NW, England.%
}

Tel: +44 01737 832164\qquad e-mail: garth.barber@virgin.net
\end{center}

\begin{center}
\textit{Abstract}
\end{center}

The theory of Self Creation Cosmology is described and found to be as concordant as the 
standard cosmological 'concordance model' with local experiments and cosmological 
observations. However it does not require the speculative hypotheses of inflation, dark 
matter or dark energy. The theory is highly predictive and when its only free parameter, H, 
is fixed empirically, all other cosmological constraints are then determined and found to be 
consistent with present observations. It is highly testable and challenges General Relativity 
in the geodetic precession measurement to be made by the Gravity Probe B satellite. The new 
theory predicts a N-S geodetic plus Thomas precession about a direction perpendicular to the 
plane of the orbit of 4.4096 arcsec/yr, that is, 2/3 of that of General Relativity. The 
predictions of the 'frame dragging' precession of that experiment are equal in both theories 
that is 0.0409 arcsec/yr. Furthermore, there are at least two other experiments that will 
distinguish between the two theories, which should be performed at the earliest opportunity, 
if the geodetic measurement proves to be consistent with the theory. These experiments ask 
the questions, "Do photons fall at the same rate as particles?" and "Is the Casimir force 
coupled to curvature?" 

\begin{center}
\ \ 
\end{center}
\newpage
\section{Introduction}

\subsection{A Possible Challenge to the Standard Model}

By the year 2003 the standard cosmological model was generally thought to be concordant with all
observations, especially the very precise WMAP data set. The General Relativity (GR) 'Hot Big 
Bang' model, having been continually refined since the 1970's, today comprises the 
standard 'concordance model'. It is now thought to be firmly established by observational 
evidence beyond any reasonable doubt. This paper questions that certitude.

It is to be noted that the standard model does indeed fit the data, but only because of several 
additions made since the mid 1970's. At that time the GR model was found to contain the 
well-known density, smoothness and horizon problems. These were resolved by the addition 
of the hypothesis of inflation, which arose out of the Higgs field, a quantum mechanical 
approach to the origin of inertial mass. 

This brought with it the additional problem that subsequent observations of galaxy clustering 
and gravitational lensing seemed to indicate that the value of the density parameter was only 
about one third whereas inflation required unity [1]; thus the initial inflation 
theory did not seem to fit the data.

Moreover, primordial cosmic abundances in the standard Big Bang scenario constrained the 
baryon density to only about 4 per cent of the critical density. Therefore it seemed that about 
one quarter of the universe's mass had not been accounted for. The additional supposition 
was thus made that this consisted of non-baryonic 'dark matter' of unknown identity. 

Subsequent observations of Type IA Supernovae, [2], [3], as well as concordance with other 
observations, including the microwave background and galaxy power spectra, implied that the 
universe was accelerating. 

Furthermore, an analysis of the anisotropies of the Cosmic Microwave Background (CMB) 
obtained from the WMAP data was seen to be consistent with a spatially flat 
universe. When interpreted within a GR environment this 'flatness' indicated that the total 
density of the universe was equal to, or very close to, the critical density, thereby 
apparently resolving the observed density disagreement with inflation. 

Therefore it seemed that about another two thirds of the universe's mass had not been 
accounted for. Again an additional supposition was made, this time that this component 
consisted of 'dark energy', also of unknown identity but having an 'anti-gravity' effect, 
in that it made cosmic expansion accelerate, thus apparently resolving the distant 
Type 1a supernovae data. There have been many unverified suggestions as to its nature such as: 
a dynamical cosmological 'constant', quintessence, decaying vacuum energy, or 
gravitational leakage from extra dimensions. 

In the present epoch the required densities of both dark matter and dark energy are 
approximately equal and, to within an order of magnitude, they approximate the 
baryonic density. There appears to be no explanation for these perplexing coincidences, 
except perhaps an anthropic argument, but this was exactly the type of improbability that 
inflation was meant to correct. Nevertheless, such has been the effectiveness 
of these additions to the basic Big Bang theory that it is now generally considered 
that the detailed observational verification of the present model consisting of 
23 per cent dark matter, 73 per cent dark energy and just 4 per cent ordinary 
matter has been robustly established.

However it may be prescient to ask, "Is not the inclusion into the standard cosmological 
concordant model of first inflation, then dark matter, and now dark energy, 
a modern example of adding 'extra epicycles', in a manner analogous to the ancient 
Ptolemaic system?" 

The force of this question is, of course, dependent on the existence of a viable alternative 
theory that does not require the continual addition of extra hypotheses. It is the intention of 
this paper to argue that such an alternative does indeed exist; it is Self Creation 
Cosmology (SCC).

\section{ The Principles of SCC}

The original SCC paper [4] was published in 1982. In that paper cosmologies were 
explored in which the matter field might be created out of self-contained gravitational and 
scalar fields. Two theories were postulated; the first was rejected on the grounds of 
non-concordant experimental violation of the Equivalence Principle, and the second was an early 
version of the present theory. That paper has generated some interest over the last twenty years 
and has been discussed in over forty-five citations. (see references [5] - [52])

The latest version of the theory, which will simply be referred to henceforth as SCC, can be 
introduced by remembering that Einstein gave some consideration to two concepts that are not 
fully included in GR, these are the local conservation of energy and Mach's Principle. At various 
times since the publication of Einstein's GR papers these concepts have been considered 
independently, in SCC they are considered together.

The first non-GR concept, which is the Local Conservation of Energy, can be appreciated by 
considering the conservation of four-momentum, $P^\nu $, of a projectile in free fall, which 
is a fundamental property of any metric theory such as GR as it necessarily follows from the 
Equivalence Principle. As a consequence the energy or 'relativistic mass' of a particle, ($P^0$), 
is not conserved, except when measured in a co-moving frame of reference, or in the Special 
Relativity (SR) limit. In any metric theory a particle's rest mass is necessarily invariant as 
it is mathematically identical to the norm of the four-momentum vector. This requirement in the 
present SCC defines the Einstein frame (EF) of the theory. The local non-conservation of 
energy is a consequence of the fact that energy is not a manifestly covariant concept, 
that is, its value is relative to the inertial frame of reference in which it is measured. 
As the Equivalence Principle does not allow a preferred frame there is no definitive 
value for energy in any metric theory.

The second non-(fully)GR concept is Mach's Principle. This suggests that inertial frames of 
reference should be coupled to the distribution of mass and energy in the universe at large, 
hence one would actually expect there to be a preferred frame, that is, a frame in which the 
universe as a whole might be said to be at rest, in which $P^0$ is conserved, in apparent
contradiction to the spirit of the Equivalence Principle. In fact such a frame of reference does 
appear to exist, it is that in which the Cosmic Background Radiation (CBR) is globally isotropic.

These two problems are linked and resolved together in the new SCC theory by the proposal that 
energy is to be locally conserved when measured in a particular, preferred, frame of 
reference as selected by Mach's principle, that is that of the Centre of Mass (CoM) 
of the system. This proposal defines what is called the Jordan (energy) Frame [JF(E)] of SCC,
in which rest mass is required to include gravitational potential energy, as 
defined in that CoM frame of reference. 

The requirement to locally conserve energy in such a Jordan Frame thus selects the 
CoM frame of reference as a "preferred foliation of space-time", to use Butterfield and 
Isham's expression [53]. Therefore, and although the question is not explored in this 
paper, in the future it might be pertinent to investigate whether such a preferred 
reference frame provides any insight into the problems at the gravitation and 
quantum theory interface.

SCC is an adaptation of the Brans Dicke (BD) theory in which the conservation requirement has 
been relaxed to allow the scalar field to interact with matter. The scalar field that 
determines inertial mass, $\phi \approx \frac 1{G_N}$ , is coupled to the large scale 
distribution of matter in motion, described by the BD field equation, [54], which takes the 
simplest general covariant form:
\begin{equation}
\Box \phi =4\pi \lambda T_M^{\;}\text{ ,}  \label{eq2}
\end{equation}
$T_{M\;}^{\;}$ is the trace, ($T_{M\;\sigma }^{\;\;\sigma }$), of the energy
momentum tensor describing all non-gravitational and non-scalar field energy.
In this theory the Brans Dicke parameter $\lambda $ is determined to be unity. [10]

The gravitational field equations are modified to explicitly include Mach's principle, following 
BD, [53], by including the energy-momentum tensor of a scalar field energy 
$T_{\phi\,\mu \nu }$ 
\begin{equation}
R_{\mu \nu }-\frac 12g_{\mu \nu }R=\frac{8\pi }\phi \left[ T_{M\mu \nu
}+T_{\phi \,\mu \nu }\right] \text{ .}  \label{eq4}
\end{equation}
where $T_{M\mu \nu }$ is the energy momentum tensors describing the matter
field. 

In SCC the conservation requirement is relaxed to allow mass to be created out of the 
gravitational and scalar fields according to the Principle of Mutual Interaction (PMI), in which 
the scalar field is a source for the matter-energy field if and only if the matter-energy field 
is a source for the scalar field.

\begin{equation}
\nabla _\mu T_{M\;\nu }^{.\;\mu }=f_\nu \left( \phi \right) \Box \phi =4\pi
f_\nu \left( \phi \right) T_{M\;}^{\;}\text{ .}  \label{eq7}
\end{equation}
As a consequence photons still do traverse null-geodesics\textit{, } at least 
\textit{in vacuo,} \textit{\ } 
\begin{equation}
\nabla _\mu T_{em\quad \nu }^{\quad \mu }=4\pi f_\nu \left( \phi \right)
T_{em}^{\;}=4\pi f_\nu \left( \phi \right) \left( 3p_{em}-\rho _{em}\right)
=0  \label{eq8}
\end{equation}
where $p_{em}$ and $\rho _{em}$ are the pressure and density of an
electromagnetic radiation field with an energy momentum tensor $T_{em\,\mu
\nu }$ , in which $p_{em}=\frac 13\rho _{em}$ and where it has been shown, [7], [10], that
\begin{equation}
f_\nu \left( \phi \right) =\frac 1{8\pi \phi }\nabla _\nu \phi \text{ .}
\label{eq7a}
\end{equation}

SCC can be thought of as a semi-metric theory in which the BD theory is adapted to include the 
local conservation of energy. It is described as 'semi-metric' as although particles do not obey 
the Equivalence Principle, photons still do. This local conservation of energy requires the 
energy expended in lifting an object against a gravitational field to be translated into an 
increase in rest mass. If $\Phi _N\left( x^\mu \right) $ is the dimensionless Newtonian 
gravitational potential defined by a measurement of acceleration in a local experiment in a frame 
of reference co-moving with the CoM, 
\begin{equation}
\frac{d^2r}{dt^2}=-\nabla \Phi _N\left( r\right) \text{ }  \label{eq105}
\end{equation}
and normalized so that $\Phi _N\left( \infty \right) $ $=0$ , then the local
conservation of energy requires (with $c = 1$)
\begin{equation}
\frac 1{m_p\left( x^\mu \right) }\mathbf{\nabla }m_p\left( x^\mu \right) =%
\mathbf{\nabla }\Phi _N\left( x^\mu \right) \text{ ,}  \label{eq66}
\end{equation}
where $m_p(x^\mu )$ is measured locally at $x^\mu $. This has the solution 
\begin{equation}
m_p(x^\mu )=m_0\exp [\Phi _N\left( x^\mu \right) ]\text{ ,}  \label{eq24}
\end{equation}
\[
\text{where}\qquad m_p(r)\rightarrow m_0\qquad \text{as\qquad }r\rightarrow
\infty \text{ .} 
\]

\subsection{The SCC Conformal Transformation}

The Jordan Frame (JF) of SCC requires mass creation, ( $\nabla _\mu T_{M\;\nu }^{\;\mu
}\neq 0$ ), therefore the scalar field is non-minimally connected to matter.
The JF Lagrangian density is, 
\begin{equation}
L^{SCC}[g,\phi ]=\frac{\sqrt{-g}}{16\pi }\left( \phi R-\frac \omega \phi
g^{\mu \nu }\nabla _\mu \phi \nabla _\nu \phi \right)
+L_{matter}^{SCC}[g,\phi ]\text{ ,}  \label{eq24a}
\end{equation}
In a general conformal transformation $%
\widetilde{g}_{\mu \nu }=\Omega ^2g_{\mu \nu }$ , mass 
is transformed according to 
\begin{equation}
m\left( x^\mu \right) =\Omega \widetilde{m}_0  \label{eq33}
\end{equation}
where $m\left( x^\mu \right) $ is the mass of a
fundamental particle in the JF and $\widetilde{m}_0$ its invariant mass in
the EF. Therefore the local conservation of energy in the SCC JF, Equations 
\ref{eq24} and \ref{eq33}, require 
\begin{equation}
\Omega =\exp \left[ \Phi _N\left( x^\mu \right) \right] \text{ .}
\label{eq34}
\end{equation}

Now, the conformal dual of the Lagrangian density, Equation \ref{eq24a}, [55], [56], by such a 
general transformation is 
\begin{eqnarray}
L^{SCC}[\widetilde{g},\widetilde{\phi }] &=&\frac{\sqrt{-\widetilde{g}}}{%
16\pi }\left[ \widetilde{\phi }\widetilde{R}+6\widetilde{\phi }\widetilde{%
\Box }\ln \Omega \right] +\widetilde{L}_{matter}^{SCC}[\widetilde{g},%
\widetilde{\phi }]  \label{eq24b} \\
&&\ -\frac{\sqrt{-\widetilde{g}}}{16\pi }\left[ 2\left( 2\omega +3\right) 
\frac{\widetilde{g}^{\mu \nu }\widetilde{\nabla }_\mu \Omega \widetilde{%
\nabla }_\nu \Omega }{\Omega ^2}+4\omega \frac{\widetilde{g}^{\mu \nu }%
\widetilde{\nabla }_\mu \Omega \widetilde{\nabla }_\nu \widetilde{\phi }}%
\Omega +\omega \frac{\widetilde{g}^{\mu \nu }\widetilde{\nabla }_\mu 
\widetilde{\phi }\widetilde{\nabla }_\nu \widetilde{\phi }}{\widetilde{\phi }%
}\right] \text{ .}  \nonumber
\end{eqnarray}

The transformation of $%
\widetilde{\phi }$ is treated in the earlier paper [7], where it is shown that $Gm$ 
is invariant and, as in the EF $%
 \widetilde{m}_0$ is defined to be invariant $%
 \widetilde{G}$, and hence $%
\widetilde{\phi }$ , are therefore constant. 
Thus in this frame $%
\widetilde{\nabla }_\nu\widetilde{\phi } =0$.
Hence if the following conditions are met: $\omega =-\frac 32$ and 
$\widetilde{\Box }\ln \Omega =0$, Equation \ref{eq24b} reduces to the canonical GR Langrangian 
density:
\begin{equation}
L^{SCC}[\widetilde{g}]=\frac{\sqrt{-\widetilde{g}}}{16\pi G_N}\widetilde{R}+%
\widetilde{L}_{matter}^{SCC}[\widetilde{g}]\text{ ,}  \label{eq24c}
\end{equation}
where matter is now minimally connected. In fact the value $\omega =-\frac 32$ has not been
arbitrarily chosen, but it is determined from the first principles of the theory [10]. 
Furthermore, the condition, $\widetilde{\Box }\ln \Omega =0$, is the vacuum condition, $%
\widetilde{\Box }\widetilde{\Phi }_N\left( \widetilde{x}^\mu \right) =0$, 
as this reduces to $\widetilde{\nabla }^2\widetilde{\Phi }_N\left( \widetilde{x}%
^\mu \right) =0$ in a harmonic coordinate system. Therefore, in this theory the conformal 
transformation of the Jordan Frame, Equation \ref{eq24a}, into the Einstein 
Frame, Equation \ref{eq24c}, results in canonical GR \textit{in vacuo}. As energy is locally 
conserved in this Jordan Frame it has been given the specific designation JF(E). 

As a result of the SCC conformal equivalence with canonical GR \textit{in vacuo}, SCC test 
particles follow the geodesics of GR in solar system experiments. Consequentially the SCC 
predictions for all the experiments tested to date are equal to those of GR.

These SCC principles have the consequence that in the Jordan Frame, in which energy is locally 
conserved in the Centre of Mass frame of reference, a photon has constant frequency and its 
energy is conserved even when passing through a gravitational field, [7]. Gravitational red shift 
is interpreted as a gain of potential energy, and hence mass, of the measuring apparatus, rather 
than the loss of (potential) energy by the photon.

\section{The Standard Formulae of SCC}

\subsection{The SCC Field Equations}

The SCC Action Principle gives rise to the following set of equations, [7]:

The scalar field equation 
\begin{equation}
\Box \phi =4\pi T_M^{\;}\text{ ,}  \label{eq143}
\end{equation}

The gravitational field equation

\begin{eqnarray}
R_{\mu \nu }-\frac 12g_{\mu \nu }R &=&\frac{8\pi }\phi T_{M\mu \nu }-\frac
3{2\phi ^2}\left( \nabla _\mu \phi \nabla _\nu \phi -\frac 12g_{\mu \nu
}g^{\alpha \beta }\nabla _\alpha \phi \nabla _\beta \phi \right)
\label{eq143a} \\
&&\ \ +\frac 1\phi \left( \nabla _\mu \nabla _\nu \phi -g_{\mu \nu }\Box
\phi \right) \text{ ,}  \nonumber
\end{eqnarray}

Finally, the creation equation, which replaces the conservation equation

\begin{equation}
\nabla _\mu T_{M\;\nu }^{\;\mu }=\frac 1{8\pi }\frac 1\phi \nabla _\nu \phi
\Box \phi \text{ .}  \label{eq144}
\end{equation}
	These field equations are manifestly covariant, there is no preferred frame of reference
or absolute time. However in order to solve them one has to adopt a specific coordinate 
system; the CoM of the system in the spherically symmetric One Body Case, or that of the 
co-moving fluid of the cosmological solution.  In those frames of reference there is a specific 
coordinate time as in the standard GR solutions.
    These JF(E) solutions of SCC, moreover, have the property not only of being in the local 
Machian frame of reference but also of locally conserving mass-energy.

\subsection{The Spherically Symmetric Solution}

The Robertson parameters are, [7],
\begin{equation}
\alpha _r=1\qquad \beta _r=1\qquad \gamma _r=\frac 13\text{ ,}  \label{eq145}
\end{equation}
and therefore the standard form of the Schwarzschild metric is 
\begin{eqnarray}
d\tau ^2 &=&\left( 1-\frac{3G_NM}r+..\right) dt^2-\left( 1+\frac{G_NM}%
r+..\right) dr^2  \label{eq146} \\
&&\ \ \ \ \ \ \ -r^2d\theta ^2-r^2\sin ^2\theta d\varphi ^2\text{ .} 
\nonumber
\end{eqnarray}
The formula for $\phi $ is 
\begin{equation}
\phi =G_N^{-1}\exp (-\Phi _N)  \label{eq108}
\end{equation}
and that for $m$ is, (Equation \ref{eq24}), 
\[
m_p\left( x_\mu \right) =m_0\exp (\Phi _N)\text{.}  
\]

\subsection{Local Consequences of the Theory}

The violation of the Equivalence Principle manifests itself as an extra force that acts on 
particles, but not photons, which acts in a similar way to gravitational force, but of one third 
the strength in the opposite direction, and which is conflated with it. Therefore there are two 
Gravitational constants, $G_N$, which applies to particles and measurable in Cavendish type 
experiments as the standard Newtonian constant and $G_m$, which applies to photons and is that 
constant that determines the curvature of space-time. These two constants relate together, [7], 
according to:
\begin{equation}
G_N=\frac 23G_m\text{ .}  \label{eq149}
\end{equation}
Hence, if normal Newtonian gravitational acceleration is $g$, the
acceleration of a massive body caused by the curvature of space-time is $%
\frac 32g$ 'downward' compensated by an 'upward' acceleration caused by the
scalar field of $\frac 12g$.

Finally in the JF(E) the radial inward acceleration of a freely falling body
is given by the non-Newtonian expression 
\begin{equation}
\frac{d^2r}{dt^2}=-\left\{ 1-\frac{G_NM}r+...\right\} \frac{G_NM}{r^2}\text{
.}  \label{eq150}
\end{equation}
In the earlier paper it was seen that the effect of this non-Newtonian
perturbation was to compensate for the effect of the scalar field upon the
curvature of space-time.

The acceleration experienced by a freely falling particle is given by 
\begin{equation}
m_0\frac{d^2r}{dt^2}=-m(r)\frac{G_NM}{r^2}\text{ .}  \label{eq151}
\end{equation}
We see that $m_0$ can be thought of as 'inertial-mass', which measures
inertia and $m(r)$ as 'gravitational mass', which interacts with the
gravitational field with 
\[
\stackunder{r\rightarrow \infty }{Lim}\text{ }m(r)=m_{0\text{ }}\text{.} 
\]

As described in the original paper, [7], the conformal equivalence between the JF(E) of SCC and 
canonical GR, results in the predictions of SCC in the standard tests being identical with GR. 
It was seen in detail in that paper that in the JF(E) of the theory, the action of the 
non-conservation of the energy-momentum tensor for matter results in an extra 'scalar-field' 
force, mentioned above, which acts on particles and which exactly compensates for the scalar 
field perturbation of the curvature of the space-time manifold. Nevertheless two definitive 
experiments were suggested in that paper, which examine the interaction of the photon and the 
vacuum energy fields with ordinary matter. Since then the geodetic precession measurement 
has been also recognised as another definitive experiment.

\section{The Cosmological Case}

\subsection{Deriving the General Cosmological Equations}

Using the Cosmological Principle the usual assumptions of homogeneity and
isotropy can be made to obtain the cosmological solutions to the field
equations. 

The privileged CoM frame in which physical units may be defined for any
epoch is now the ''rest frame'' for the universe as a whole. Presumably it
should be identified physically with that frame in which the microwave
background radiation is globally isotropic.

There are two questions to ask in order that a Weyl metric may be set up spanning extended 
space-time, ''What is the invariant standard by which objects are to be measured?'' and 
''How is that standard to be transmitted from event to event in order that the comparison can 
be made?'' In the SCC EF, and GR, the principle of energy-momentum conservation, i.e. invariant
rest mass, determines that standard of measurement to be fixed rulers and regular clocks. In the 
SCC JF(E), on the other hand, the principle of the local conservation of energy determines that 
standard of measurement to be a ''standard photon'', which is to be taken from the CMB in the 
cosmological case. The inverse of its frequency determines the standard of time and space 
measurement, and its energy determines the standard of mass, all defined in the CoM, Machian, 
frame of reference. According to SCC, a gravitational field, i.e. the curvature of space-time, 
is to be described in the JF(E), whereas observations using atomic apparatus, based on an 
atomic clock, are referred to the EF. The two frames have to be transformed as appropriate.

There are four equations to consider that are treated in the earlier paper, [7]. 
The first is the Gravitational Field Equation \ref{eq143a}, which is exactly the 
same as the BD equation with $%
\omega =-\frac 32$ . The second is the Scalar Field
Equation \ref{eq2}. In GR the third equation is the conservation equation,
which is replaced in SCC by the Creation Field Equation \ref{eq144}. The
fourth equation is some equation of state, such as the dust filled universe $%
p=0$, or the early radiation dominated universe in which $p=\frac 13\rho $.
The SCC field equations demand an exotic equation of state.

The two gravitational cosmological equations are 
\begin{equation}
\left( \frac{\stackrel{.}{R}}R\right) ^2+\frac k{R^2}=+\frac{8\pi \rho }{%
3\phi }-\frac{\stackrel{.}{\phi }\stackrel{.}{R}}{\phi R}-\frac 14\left( 
\frac{\stackrel{.}{\phi }}\phi \right) ^2\text{ ,}  \label{eq197}
\end{equation}
\begin{equation}
\frac{\stackrel{..}{R}}R+\left( \frac{\stackrel{.}{R}}R\right) ^2+\frac
k{R^2}=-\frac 16\left( \frac{\stackrel{..}{\phi }}\phi +3\frac{\stackrel{.}{%
\phi }\stackrel{.}{R}}{\phi R}\right) +\frac 14\left( \frac{\stackrel{.}{%
\phi }}\phi \right) ^2\text{ .}  \label{eq199}
\end{equation}

The scalar cosmological equation 
\begin{equation}
\stackrel{..}{\phi }+\,3\frac{\stackrel{.}{\phi }\stackrel{.}{R}}R=4\pi
\left( \rho -3p\right) \text{ .}  \label{eq202}
\end{equation}

The creation cosmological equation is 
\begin{equation}
\stackrel{.}{\rho }\,=-3\frac{\stackrel{.}{R}}R\left( \rho +p\right) +\frac
1{8\pi }\frac{\stackrel{.}{\phi }}\phi \left( \stackrel{..}{\phi }+\,3\frac{%
\stackrel{.}{\phi }\stackrel{.}{R}}R\right) \text{ .}  \label{eq204}
\end{equation}
(It is a moot point whether the scalar field $\phi $ is generated by the
distribution of mass and energy via Equation \ref{eq202}, or whether mass is
generated by the scalar field via Equation \ref{eq204}.)

Finally the equation of state remains 
\begin{equation}
p=\sigma \rho \text{ ,}  \label{eq205}
\end{equation}
where the equations determine $\sigma =-\frac{1}{3}$ in the SCC universe.

\subsection{The SCC Cosmological Solution}

The five independent Equations, \ref{eq197}, \ref{eq199}, \ref{eq202}, \ref
{eq204} and \ref{eq205} and the sixth relationship, provided by the
conservation of a free photon's energy in the JF(E) together with Stephan's
Law, provide a solution for the six unknowns $R(t)$, $\phi (t)$, $\rho (t)$, 
$p(t)$, $k$ and $\sigma $. There are also the boundary conditions at $t=t_0$
(present epoch), $R_0$, $\phi _0$, $\rho _0$, and $p_0$.

The cosmological 'self-creation equation' is found to be, [7], 
\begin{equation}
\rho =\rho _0\left( \frac R{R_0}\right) ^{-3\left( 1+\sigma \right) }\left(
\frac \phi {\phi _0}\right) ^{\frac 12\left( 1-3\sigma \right) }\text{ ,}
\label{eq206}
\end{equation}
which is the equivalent GR expression with the addition of the last factor
representing cosmological 'self-creation'. However for a photon gas $\sigma
=+\frac 13$ so Equation \ref{eq206} reduces to its GR equivalent, consistent
with the Principle of Mutual Interaction that there is no interaction
between a photon and the scalar field, 
\begin{equation}
\rho _{em}=\rho _{em\,0}\left( \frac R{R_0}\right) ^{-4}\text{ .}
\label{eq207}
\end{equation}
Since $\rho _{em}\propto T_{em}^4$where $T_{em}$ is the Black Body
temperature of the radiation, the GR relationship $T_{em}\propto R^{-1}$
still holds. Also as the wavelength $\lambda _{em}$ of maximum intensity of
the Black Body radiation is given by $\lambda _{em}\propto T_{em}^{-1}$, the SCC
JF(E) retains the GR relationship 
\begin{equation}
\lambda _{em}\propto R\text{ .}  \label{eq208}
\end{equation}

However in the SCC JF(E) $\lambda _{em}$ is constant for a free photon, even
over curved space-time, and it is particle masses that vary. Therefore in
the JF(E) Equation \ref{eq208} becomes simply 
\begin{equation}
R=R_0\text{ .}  \label{eq209}
\end{equation}
In the Jordan energy frame the universe is static when measured by light,
that is, as a co-expanding ''light ruler'' is unable to detect the expanding
universe there is no expansion.

The cosmological gravitational and scalar field equations are solved to
yield 
\begin{equation}
\phi =\phi _0\exp \left( H_0t\right) \text{ ,}  \label{eq214}
\end{equation}
where $H_0$ is Hubble's 'constant' in the present epoch, defined by $t=0$ ,
and $\phi _0=G_N^{-1}$. By definition $G_N$ is the value measured in
''Cavendish type'' experiments in the present epoch. Note the theory admits
a cosmological ground state solution, $g_{\mu \nu }\rightarrow \eta _{\mu
\nu }$ and $\nabla _\mu \phi =0$ only when $t\rightarrow -\infty $, that is
at the ''Big Bang'' itself. Equations \ref{eq202}, \ref{eq205}, \ref{eq209}, 
\ref{eq214} and $\sigma =-\frac{1}{3}$ yield 
\begin{equation}
\frac{8\pi \rho }{\phi _0}=H_0^2\exp \left( H_0t\right) \text{ .}
\label{eq221}
\end{equation}
This can be written in the form 
\begin{equation}
\rho =\rho _0\exp \left( H_0t\right) \qquad  \label{eq222}
\end{equation}
\begin{equation}
\text{where\ }\rho _0=\frac{H_0^2}{8\pi G_N}\text{ ,}  \label{eq238}
\end{equation}
if now, as usual, the critical density is defined $\rho _c=\frac{3H_0^2}{%
8\pi G_N}$, then $\rho _0=\frac 13\rho _c$. Hence the cosmological density
parameter $\Omega _c$ 
\begin{equation}
\Omega _c=\frac 13\text{ . }  \label{eq239}
\end{equation}
Therefore, in this theory there is no need for 'Dark Energy'. The
cosmological density parameter $\Omega _c$ comprises of baryonic (plus any
cold dark matter) and radiation (plus any hot dark matter) components
together with that of false vacuum energy. As the total pressure is
determined by the constraints of the cosmological equations $\sigma =-\frac{1}{3}$, 
together with Equation \ref{eq222}, the total cosmological pressure
is given by 
\begin{equation}
p=-\frac 13\rho _0\exp \left( H_0t\right) \text{ .}  \label{eq241}
\end{equation}
To explain this it is suggested that a component of the cosmological
pressure and density is made up of false vacuum. In other words there is a
''remnant'' vacuum energy made up of contributions of zero-point energy from
every mode of every quantum field that would have a natural energy
''cut-off'' $E_{\max }$, which in the cosmological case is determined, and
limited, by the solution to the cosmological equations. If the total density comprises of
both a baryon density together with a false vacuum density then the cosmological equations
require 
\begin{equation}
\rho _b=2\rho _f\text{ .}  \label{eq250}
\end{equation}
Therefore the density parameter for cold matter (visible and dark) is 
\begin{equation}
\Omega _b=\frac 29\approx 0.22\ \text{.}  \label{eq251}
\end{equation}

Assuming baryon conservation in a static universe, the inertial mass of a
fundamental particle must be given by 
\begin{equation}
m_i=m_0\exp \left( H_0t\right) \text{ .}  \label{eq223}
\end{equation}

\subsection{The Transformation Into the Einstein Frame (EF)}

Measurements of curvature or the wavelength/energy of a photon are made in
the JF(E), however the physics of atomic structures is naturally described
in the EF. It is now necessary to transform the units used in the JF(E) into
the system used in physical measurement using atomic apparatus; that is the
EF. The two frames are conformally related, where $%
\Omega $ is the parameter of conformal transformation, 
\[
g_{\mu \nu }\rightarrow \widetilde{g}_{\mu \nu }=\Omega ^2g_{\mu \nu }\text{
,} 
\]
where the interval is invariant under the transformation

\begin{equation}
d\tau ^2=-g_{\mu \nu }dx^\mu dx^\nu =-\widetilde{g}_{\mu \nu }d\widetilde{x}%
^\mu d\widetilde{x}^\nu \text{ . }  \label{eq224}
\end{equation}

Now mass transforms according to Equation \ref{eq33}

\[
m\left( x^\mu \right) =\Omega \widetilde{m}\text{ ,} 
\]
therefore Equation \ref{eq223} requires in the cosmological solution to the
field equations 
\begin{equation}
\Omega =\exp \left( H_0t\right) \text{ .}  \label{eq225}
\end{equation}
From which the transform of length and time is obtained by integrating along space-like and
time-like paths respectively, 
\begin{equation}
\widetilde{L}=L_0\exp \left( H_0t\right) \text{ }  \label{eq226}
\end{equation}
\begin{equation}
\text{and\ }\triangle \widetilde{t}=\triangle t\exp \left( H_0t\right) \text{
.}  \label{eq227}
\end{equation}
These transformations are consistent with using the Bohr/Schr\"odinger/Dirac
models of an atom to measure length and time under mass transformation.

The two time scales relate to each other as follows 
\begin{equation}
\widetilde{t}=\frac 1{H_0}\exp \left( H_0t\right) \text{\qquad and \qquad }%
t=\frac 1{H_0}\ln \left( H_0\widetilde{t}\right) \text{ ,}  \label{eq227a}
\end{equation}
where $\widetilde{t}$ is time measured from the ''Big Bang'' in the EF, and $%
t$ is time measured from the present day in the JF(E).

Applying this transformation to the universe's scale factor in two steps,
the first step yields 
\begin{equation}
\widetilde{R}=R_0\exp \left( H_0t\right) \text{ .}  \label{eq228}
\end{equation}
This expression uses mixed frames, that is, length is in the EF and time is
in the JF(E). If we now substitute for $t$ in Equation \ref{eq228} we obtain
the scale factor of the universe in the EF. 
\begin{equation}
\widetilde{R}=R_0\frac{\widetilde{t}}{\widetilde{t}_0}\text{ .}
\label{eq233}
\end{equation}

Thus when measured by physical, that is atomic, rulers and clocks the universe is seen to
expand linearly from a ''Big Bang''. The deceleration parameter 
\begin{equation}
q=-\left( \frac{\stackrel{..}{\widetilde{R}}}{H^2\widetilde{R}}\right) =0%
\text{ .}  \label{eq234}
\end{equation}
Therefore the horizon, smoothness and density problems of classical GR
cosmology, which all arise from a positive, non zero $q$, do not feature in
SCC. Hence it is unnecessary to invoke Inflation in this theory and indeed,
with Equation \ref{eq228}, SCC might be considered to be a form of
''Continuous Inflation''.

The curvature constant $k$ is given by Equations \ref{eq199} and \ref{eq197} 
\begin{equation}
\frac k{R_0^2}=+\frac 1{12}H_0^2\text{ ,}  \label{eq242}
\end{equation}
so $k$ is positive definite, 
\begin{equation}
k=+1\text{ ,}  \label{eq243}
\end{equation}
that is, the universe is finite and unbounded. From Equation \ref{eq242} $R_0$
can be derived in terms of the Hubble time 
\begin{equation}
R_0=\sqrt{12}H_0^{-1}\text{ .}  \label{eq244}
\end{equation}

\section{Summary}

In the JF(E), where energy is conserved but energy-momentum is not, photons are the means 
of measuring length, time and mass. Proper mass increases with gravitational potential 
energy and as a consequence cosmological red shift is caused by a secular, exponential, 
increase of particle masses and not cosmological expansion. The universe is static, 
in which atomic rulers 'shrink' exponentially, and eternal, in which atomic clocks 
'speed up' exponentially.

In the EF, where energy-momentum is conserved and particle proper masses are invariant, atoms 
are the means of measuring length, time and mass. As the scalar field adapts the cosmological 
equations the universe expands linearly from a Big Bang, it is a "freely coasting" universe. 

\subsection{Observational consequences of the Theory}

In order to compare the theory with observations and experiments to date it is important to 
note both that, {\it in vacuo} SCC test particles follow GR geodesic trajectories and, 
although the Robertson parameters are given by Equation \ref{eq145} as:
$
\alpha _r=1 \text{ , } \beta _r=1 \text{ and } \gamma _r=\frac 13\text{ ,}  
$
so that in particular we have  
$
\gamma _r=\frac 13\ 
$, which is less than the GR equivalent, this is compensated in most observations by an 
increase in the gravitational 'constant':
$
G_m=\frac 32G_N\text{ .}  
$ 
Hence the theory of SCC is concordant with all experiments and observations to date that 
otherwise have been thought to verify GR. For example, in the Gravity Probe B satellite 
experiment the 'frame dragging' prediction is given by the expression:
\begin{equation}
\stackrel{3}{g}_{i0}=-4G_m\left( \frac {1+\gamma _r }2\right) \int \frac{%
\stackrel{1}{T}^{i0}(x^{\prime },t)}{\left| x-x^{\prime }\right| }d^3x\text{
.}  \label{eq181}
\end{equation}
so the SCC values for $
G_m=\frac 32G_N\
$ 
and 
$
\gamma _r=\frac 13\ 
$ 
give the same result as GR in which $
G_m=G_N\
$ 
and 
$
\gamma _r=1 
$ . On the other hand in the geodetic prediction the precession is given by the expression
\begin{equation}
\frac 12\left( 2\gamma _r +1\right) \frac{G_mM_{\oplus }^{}}{R^3}\mathbf{v}%
_s\times \mathbf{X}\text{ ,}  \label{eq171}
\end{equation}
which in GR, where $\gamma _r =1$ and $G_m=G_N$ , predicts a precession for the
Gravity B Probe gyroscope of 6.6144 arc sec/yr about a direction perpendicular
to the plane of the orbit. However, in SCC $\gamma _r =\frac 13$ and $%
G_m=\frac 32G_N$ , so the theory predicts a value 5/6 of this or just 5.5120
arc sec/yr. In SCC there is also a Thomas precession, which has to be subtracted from the 
geodetic precession, of 
 \begin{equation}
\frac{1}{6} \frac{G_{m}M_{\oplus }}{R^{3}}%
\mathbf{v}_{s}\times \mathbf{X}\text{ ,}  \label{172}
\end{equation}%
Therefore, the SCC theory prediction of a N-S precession of the GP-B gyroscope 
is $%
\frac{2}{3}$ of that of the GR prediction, or just 4.4096 arcsec/yr.This crucial measurement 
will be the first experiment ever that will be able to distinguish between the two theories.

The consequence of this theory is the realisation that there are two distinct ways of 
interpreting observations of the universe. In a laboratory on Earth scientific 
observations defining units of length, time and mass/energy have to be referred 
to an atomic standard. However, astrophysical and cosmological observations only 
sample photons and not particles from the depths of the universe. How then does 
the measurement of standard units made in a laboratory here and now on Earth 
relate to an event that occurred millennia ago in a distant part of the universe? 
In particular the problem is rooted in the variation of energy levels, and hence 
frequency and wavelength, of photons over and above that caused by the Doppler 
effect, because of gravitational and cosmological red shift. 

Based on the Equivalence Principle, GR defines the proper rest mass of a particle 
to be invariant, therefore that theory requires the measure of standard 
units to be atomic 'rigid' rulers and atomic 'regular' clocks. However, in 
doing so it violates the conservation of mass-energy as described in the 
Introduction section above.

On the other hand, if a gravitational theory were to include the local conservation 
of energy, as in the theory of SCC, then an atom's rest mass would vary with 
gravitational potential energy, whereas a photon's energy would be decoupled 
from the effects of curvature. If this indeed occurs then a choice may be made 
as to the invariant standard by which units of length, time and mass/energy 
are measured. This choice of the unit used for comparison is between a 'standard' 
atom, taken from a laboratory, or a 'standard' photon, sampled from the CMB. 
Observations of the cosmos would then fall into one of two complementary interpretations: 
either that of the Jordan Frame static universe, which is eternal with no origin in time, 
or that of the Einstein Frame strictly linearly expanding universe, which has had 
an 'origin' in a 'Big Bang' at one 'Hubble Time' in the past. Either model would be a 
valid interpretation of the data, the JF(E) would be the appropriate frame to observe 
gravitational orbits and the curvature of space-time, and the EF would be the 
appropriate frame to observe atomic processes such as primordial nucleosynthesis.

It is remarkable that both these models, the static universe and the freely coasting universe, 
have already been independently investigated and both have been found to be surprisingly 
concordant with accepted cosmological constraints, including Big Bang nucleosynthesis 
abundances, distant Type Ia supernovae observations and the WMAP CMB anisotropy data. 

The static universe has been investigated heuristically by Ostermann [57], and found to be 
able to fit the standard concordance model perfectly [58]. 

The strictly linearly expanding or freely coasting model has been investigated by 
Kolb, [59], Batra, Lohiya, Mahajan and Mukherjee, [60], Dev, Safonova, Jain and Lohiya, [61], 
Gehaut, Mukherjee, Mahajan and Lohiya, [62], and Gehaut, Kumar, Geetanjali and Lohiya,[63].
Their motivation in exploring such a cosmology was the recognition that the model 
would not have suffered from the original density, smoothness and horizon problems 
of the standard GR theory. The latter paper reviews their results and finds the 
freely coasting universe fits the Type 1a supernovae data. Moreover, the recombination 
history gives the location of the primary acoustic peaks of the WMAP data in the 
same range of angles as that given in standard cosmology. Safonova, [64], in her PhD 
thesis, reports that gravitational lensing is also consistent with the linearly 
expanding freely coasting model. A further remarkable result of this model is the 
analysis of nucleosynthesis in the Big Bang. They, [60], calculate that a baryon entropy 
ratio of 
$ 
\eta = 5 \times 10^{-9} $ 
yields 23.9 per cent Helium and $ 
10^{8} $ 
times the metallicity of the standard scenario, which although large is still of the 
same order of magnitude as seen in the lowest metallicity objects. Therefore, one 
prediction of the theory is that a significant proportion of intergalactic medium 
metallicity should be primordial. 

A further consequence is, interestingly, that the production of this amount of helium 
requires a baryon density parameter of about 0.2. As the total non-false vacuum energy density 
is required by SCC to be only 0.22, there is no need for unknown dark matter. In SCC, this 
component of the cosmic density parameter is in the form of intergalactic cold baryonic matter. 

Furthermore the cosmological solution requires the universe to have an overall density 
parameter of only one third, yet be closed and conformally spatially flat. Hence the 
theory does not require dark energy, nor a significant amount of dark matter, to account 
for the present cosmological constraints.

Finally as described in a former e-print, [8], a 'time-slip' exists in SCC between 
atomic 'clock' time on one hand and gravitational ephemeris and cosmological time on the other, 
which would result in an apparent sunwards acceleration of the Pioneer spacecraft as 
indeed is observed, [67], [68], [69] and [57].

There are three experiments that are able to distinguish between the two theories, one of which 
is the Gravity Probe B geodetic measurement described above. The other two are described in 
references [7] and [9]. These ask the questions, "Do photons fall at the same rate as 
particles?" and "Is there a cut-off to the Casimir force that approaches zero as curvature 
approaches flatness detectable in the solar system somewhere beyond the orbit of Jupiter?" 

The first of these experiments might consist of an annulus of tiny mirrors designed to reflect 
one half of a split beam along a path length of about two kilometres while the other half is 
simply reflected once along a short path length before the two half beams are 
recombined in an interferometer. Inverting the apparatus in Earth orbit would produce a 
shift, or not, in the interference pattern depending on whether the photon beam is falling at 
a different, or the same rate as the apparatus. This experiment might be a suitable addition 
to the STEP (Satellite Test of the Equivalence Principle) programme, [70], either as an extra 
component of the planned spacecraft, or as a separate experiment possibly carried in the 
Space Shuttle.

The second experiment would measure the Casimir force between two close plates at an 
increasing range from the Sun and other large planets in order to detect a cut-off that 
depended on the Sun's gravitational field. It would be a suitable addition to the 
"yo-yo" craft concept, suggested by Michael Martin Nieto and Slava G. Turyshev, [71], 
in order to test the Pioneer anomaly.

It is suggested that these experiments be performed at the earliest opportunity if the 
geodetic measurement should indeed prove to be consistent with the Self Creation theory. 

\section{Conclusion}

In conclusion SCC does not require inflation to resolve the density, smoothness and horizon 
problems, as they do not exist in the theory; it does not require dark matter as the mass density 
of baryonic matter is determined to be 0.22, and it does not require dark energy as the model is
conformally flat with a total density of 0.33. It does require a false vacuum energy density, 
which is required by the cosmological equations to be 0.11, thus resolving the 'lambda' problem. 
It is therefore a testable theory, which is as concordant with the cosmological constraints 
as the standard GR model, but without these additional hypotheses, and it should be 
considered as a valid alternative to that theory.


\section{References}

\begin{thebibliography}{set}
\bibitem{1}
Chae, K.-H., \& Biggs, A.D., \& Blandford, R.D., \& Browne, I.W.A., \& de Bruyn, A.G., 
\& Fassnacht, C.D., \& Helbig, P., \& Jackson, N.J., \& King, L.G., \& Koopmans, L.V.E.,  
\& Mao, S., \& Marlow, D.R., \& McKean, J.P., \& Myers, S.T. \& Norbury, M., \& Pearson, T.J., 
\& Phillips, P.M., \& Readhead, A.C.S., \& Rusin, D., \& Sykes, C.M., \& Wilkinson, P.N., 
\& Xanthopoulos, E., \& York, T. : 2002, Constraints on Cosmological Parameters from the 
Analysis of the Cosmic Lens All Sky Survey Radio-Selected Gravitational Lens Statistics. 
{\it astro-ph/0209602}
\bibitem{2}
Perlmutter, S., \& Aldering, G., \& Goldhaber, G., \& Knop, R. A., \& Nugent, P., \& Castro, P. G., 
\& Deustua, S., \& Fabbro, S., \& Goobar, A., \& Groom, D. E., \& Hook, I. M., \& Kim, A. G.,
\& Kim, M. Y., \& Lee, J. C., \& Nunes, N. J., \& Pain, R., \& Pennypacker, C. R., \& Quimby, R.,
\& Lidman, C., \& Ellis, R. S., \& Irwin, M., \& McMahon, R. G., \& Ruiz-Lapuente, P., \& Walton, N.,
\& Schaefer, B., \& Boyle, B. J., \& Filippenko, A. V., \& Matheson, T., \& Fruchter, A. S., 
\& Panagia, N., \& Panagia, N., \& Newberg, H. J. M., \& Couch, W. J., \& Newberg, H. J. M.; 
Couch, W. J.: 1999, {\it Ap. J.} 517, 565. Measurements of Omega and Lambda from 42 High-Redshift 
Supernovae
\bibitem{3}
A. G. Riess et al., Astron.J. 116 (1998) 1009-1038, astro-ph/9805201
Observational Evidence from Supernovae for an Accelerating Universe and a Cosmological Constant
\bibitem{4}
Barber, G.A. : 1982, {\it Gen Relativ Gravit.} 14, 117. On Two Self Creation
Cosmologies.
\bibitem{5}
Abdel-Rahman, A.M.M. :1992, {\it Astrophysics and Space Science} 189, 1.
Singularity-free self-creation cosmology.
\bibitem{6}
Abdussattar \& Vishwakarma, R. G. : 1997, {\it Classical and Quantum Gravity} 14 pp 945-953
Some FRW models with variable G and $\Lambda$.
\bibitem{7}
Barber, G.A. : 2002a, {\it Astrophysics and Space Science} 282, 4, pp 683-731. A
New Self Creation Cosmology.
\bibitem{8}
Barber, G.A. : 2002b, {\it arXiv:gr-qc/0212111} The Principles of Self Creation Cosmology 
and its comparison with General Relativity.
\bibitem{9}
Barber, G.A. : 2003a, {\it arXiv:gr-qc/0302026} Experimental tests of the new Self 
Creation Cosmology and a heterodox prediction for Gravity Probe B.
\bibitem{10}
Barber, G.A. : 2003b, {\it arXiv:gr-qc/0302088} The derivation of the coupling constant in the
new Self Creation Cosmology.
\bibitem{11}
Brans, C.H. :1987, {\it Gen Relativ Gravit.}19, 949. Consistency of field
equations in self-creation cosmologies.
\bibitem{12}
Maharaj, S.D. \& Beesham, A. : 1988, {\it Astrophysics and Space Science} 140, 1.
The vacuum Friedmann-type solutions in Barber's theory of gravitation.
\bibitem{13}
Mohanty, G. \& Mishra, B. : 2002, {\it Astrophysics and Space Science} 281, 3.
Vacuum cosmological models in Einstein and Barber theories.
\bibitem{14}
Mohanty, G. \& Mishra, B. \& Biswal, A.K. : 2002, {\it Czechoslovak Journal of Physics} 52, 12. pp 1289-1296
Cosmological Models in Barber's Second Self-Creation Theory
\bibitem{15}
Mohanty, G, Sahu, R.C. \& Panigrahi, U.K. : 2002, {\it Astrophysics and Space
Science} 281, 3. Exact Bianchi Type-I cosmological model in modified theory of General Relativity.
\bibitem{16}
Mohanty, G, Sahu, R.C. \& Panigrahi, U.K. : 2003, {\it Astrophysics and Space
Science} 284, 3. Micro and Macro Cosmological Model in Barber's Second Self-Creation Theory.
\bibitem{17}
Luis O. Pimentel, {\it Astrophys. Space Sci.} {\bf116}, 395 (1985). Exact
self-creation cosmological solutions.
\bibitem{18}
Pradhan A. \& Pandey H.R. : 2002, {\it International Journal of Modern Physics D}, {\it arXiv:gr-qc/0207027} v1 4 Jul 2002.
Bulk Viscous Cosmological Models in Barber's Second Self Creation Theory
\bibitem{19}
Pradhan A. \& Vishwakarma, A. K. : 2002, {\it International Journal of Modern
Physics D}, Vol. 11, No. 8, 1195-1207. LRS Bianchi Type-I Cosmological Models
in Barber's Second Self Creation Theory.
\bibitem{20}
Rahman, A.M.M. \& Adbel. : 1993, {\it Astrophysics and Space Science} 189, 1.
Singularity-free Self-Creation Cosmology.
\bibitem{21}
Ram, S. \& Singh, CP. : 1998a, {\it Astrophysics and Space Science} 257, 1. Early
universe in self-creation cosmology.
\bibitem{22}
Ram, S. \& Singh, CP. : 1998b, {\it Astrophysics and Space Science} 257, 2.
Anisotropic bianchi type-II cosmological models in self-creation cosmology.
\bibitem{23}
Reddy, D.R.K. : 1987a, {\it Astrophysics and Space Science} 132, 2. Vacuum model
in self-creation cosmology.
\bibitem{24}
Reddy, D.R.K. : 1987b, {\it Astrophysics and Space Science} 132, 2. Bianchi type-I
vacuum model in self-creation cosmology.
\bibitem{25}
Reddy, D.R.K. : 1987c, {\it Astrophysics and Space Science} 133, 1. Vacuum
Friedmann model in self-creation cosmology.
\bibitem{26}
Reddy, D.R.K. : 1987d, {\it Astrophysics and Space Science} 133, 2. Bianchi type-I
universe filled with disordered radiation in self-creation cosmology.
\bibitem{27}
Reddy, D.R.K. : 1988, {\it Astrophysics and Space Science} 140, 1. Birkhoff's theorem in a
conformally-invariant scalar field theory
\bibitem{28}
Reddy, D.R.K., Avadhanulu, M.B. \& Venkateswarlu, R. : 1987, {\it Astrophysics 
and Space Science} 134, 1. Birkhoff-type theorem for electromagnetic-fields
in self-creation cosmology
\bibitem{29}
Reddy, D.R.K., Avadhanulu, M.B. \& Venkateswarlu, R. : 1988 {\it Astrophysics 
and Space Science} 141, 1. A static conformally-flat vacuum model in
self-creation cosmology.
\bibitem{30}
Reddy, D.R.K. \& Venkateswarlu, R. : 1988, {\it Astrophysics and Space Science}
147, 1. Nonexistence of static conformally-flat solutions in self-creation
cosmology.
\bibitem{31}
Reddy, D.R.K. \& Venkateswarlu, R. : 1989, {\it Astrophysics and Space Science}
155, 1. Bianchi type-VI models in self-creation cosmology.
\bibitem{32}
Sahu, R.C. \& Panigrahi, U.K. :2003, {\it Astrophysics and Space Science} 288(4): 601-610.
Bianchi Type-1 vacuum models in modified theory of general relativity.
\bibitem{33}
Sanyasiraju, Y.V.S.S. \& Rao, V.U.M. : 1992, {\it Astrophysics and Space Science}
189, 1. Exact bianchi-type-VIII and Bianchi-type-IX models in the presence
of the self-creation theory of cosmology.
\bibitem{34}
Shanthi, K. \& Rao, V.U.M. : 1991, {\it Astrophysics and Space Science} 179, 1.
Bianchi type-II and type-III models in self-creation cosmology.
\bibitem{35}
Singh, T., Singh, T. \& Srivastava, O.P. : 1987, {\it International Journal of
Theoretical Physics} 26, 9. Birkhoff theorem in self-creation cosmology.
\bibitem{36}
Singh, T. \& Singh, T. :1984, {\it Astrophysics and Space Science} 102, 1. Some
general results in self-creation cosmologies.
\bibitem{37}
Singh, T. : 1986, {\it Journal of Mathematical Sciences} 27, 4. Birkhoff-type
theorem in self-creation cosmology.
\bibitem{38}
Singh, T. : 1989, {\it Astrophysics and Space Science} 152, 1. Static vacuum
fields in self-creation cosmology.
\bibitem{39}
Soleng, Harald H. : 1987a, {\it Astrophysics and Space Science} 138, 1. A note on
vacuum self-creation cosmological models.
\bibitem{40}
Soleng, Harald H. : 1987b, {\it Astrophysics and Space Science} 139, 1.
Self-creation cosmological solutions.
\bibitem{41}
Venkateswarlu, R. \& Reddy, D.R.K. : 1988, {\it Astrophysics and Space Science} 150, 2. 
Plane-symmetric vacuum in self-creation cosmology.
\bibitem{42}
Venkateswarlu, R. \& Reddy, D.R.K. : 1989a, {\it Astrophysics and Space Science} 151, 1. 
Vacuum Friedmann models in self-creation cosmology.
\bibitem{43}
Venkateswarlu, R. \& Reddy, D.R.K. : 1989b, {\it Astrophysics and Space Science} 151, 2. 
Bianchi type-V radiating model in self-creation cosmology.
\bibitem{44}
Venkateswarlu, R. \& Reddy, : 1989c, {\it Astrophysics and Space Science} 152, 2.
An anisotropic cosmological model in self-creation cosmology.
\bibitem{45}
Venkateswarlu, R. \& Reddy, : 1989d, {\it Astrophysics and Space Science} 161, pg125. 
Vacuum Bianchi type V and VI(0) cosmological models in a new scalar-tensor theory of gravitation.
\bibitem{46}
Venkateswarlu, R. \& Reddy, : 1989e, {\it Astrophysics and Space Science} 159, pg. 173.
On Birkhoff's theorem in Bergmann-Wagoner theory
\bibitem{47}
Venkateswarlu, R. \& Reddy, : 1989f, {\it Astrophysics and Space Science} 155, pg. 135. 
Bianchi type-VI(0) models in self-creation cosmology.
\bibitem{48}
Venkateswarlu, R. \& Reddy, D.R.K. : 1990, {\it Astrophysics and Space Science}
168, 2. Bianchi type-I models in self-creation theory of gravitation.
\bibitem{49}
Wolf, C.: 1986, {\it 1986PhyS...34..193W}
Non-conservative gravitation and Kaluza Klein cosmology
\bibitem{50}
Wolf, C.: 1988, {\it Astronomische Nachrichten} 309, 3 pgs.173-175. 
Higher order curvature terms in theories with creation.
\bibitem{51}
Wolf, C.: 1988, {\it 1988PhyS...38..129W}
Can inflation take place in a closed universe admitting creation?
\bibitem{52}
Wolf, C.: 1992, {\it Astronomische Nachrichten} 313, 3 pgs.133-137.
Looking for a massive KeV pseudo-scalar in gamma ray bursts.
\bibitem{53}
Butterfield, J. \& Isham, C. J. : 2001. {\it Physics meets Philosophy at the Planck Scale}, 
ed. by C. Callender and N. Huggett. Cambridge University Press.
\bibitem{54}
Brans, C.H. \& Dicke, R.H. : 1961, {\it Phys. Rev.} 124, 925.
Mach's Principle and a Relativistic Theory of Gravitation
\bibitem{55}
Dicke, R.H. : 1962, {\it Phys. Rev.} 125, 2163. 
Mach's Principle and Invariance under Transformation of Units.
\bibitem{56}
Synge, J.L. : 1955, {\it Relativity: the General Theory.} North Holland: Amsterdam. 
\bibitem{57}
Ostermann, P. : 2002, {\it arXiv:gr-qc/0212004}. 
Relativity Theory and a Real Pioneer Effect
\bibitem{58}
Ostermann, P. : 2003, {\it arXiv:astro-ph/0312655}.
The Concordance Model - a Heuristic Approach from a Stationary Universe.
\bibitem{59}
Kolb, E.W. : 1989, {\it The Astrophysical Journal} 344, 543.
A coasting cosmology.
\bibitem{60}
A. Batra, D. Lohiya, S. Mahajan, A. Mukherjee, {\it Int. J. Mod. Phys.}, D9, 757 (2000).
Nucleosynthesis in a Universe with a Linearly Evolving Scale Factor.
\bibitem{61}
Dev, A. \& Safonova, M. \& Jain, D. \& Lohiya, D. : 2002, {\it arXiv:astro-ph/0204150}.
Cosmological Tests for a Linear Coasting Cosmology.
\bibitem{62}
Gehaut, S. \& Mukherjee, A. \& Mahajan, S. \& Lohiya, D. : 2002, {\it arXiv:astro-ph/0209209}.
A "Freely Coasting" Universe.
\bibitem{63}
Gehaut, S. \& Kumar, P. \& Geetanjali. \& Lohiya, D. : 2003, {\it arXiv:astro-ph/0306448}.
A Concordant "Freely Coasting Cosmology".
\bibitem{64}
Safonova, M. : 2002. PhD Thesis, University of Delhi. 2004, {\it arXiv:astro-ph/0401542}.
Gravitational Lensing in Standard and Alternative Cosmologies.
\bibitem{65}
Etherington, J.M.H. : 1933. {\it Phil. Mag.} 15, 761 
\bibitem{66}
Anderson, J.D. \& Laing, Ph.A. \& Lau, E.L. \&  Liu, A.S. \& Nieto, M.M \& Turyshev, S.G. : 1998. 
{\it arXiv:gr-qc/0104064} {\it Phys. Rev. Lett} 81, pg.2858. 
Study of the anomalous acceleration of Pioneer 10 and 11
\bibitem{67}
Anderson, J.D. \& Laing, P.A. \& Lau, E.L. \& Liu, A.S. \& Nieto, M.M. \& Turyshev, S.G. : 2002. 
{\it Physical Review D}, 65, 8, id. 082004.
Study of the anomalous acceleration of Pioneer 10 and 11.
\bibitem{68}
Mbelek, J.P. \& Michalski, M. : 2003. {\it arXiv:gr-qc/0310088}.
Can conventional forces really explain the anomalous acceleration of Pioneer 10/11?
\bibitem{69}
Worden, P.W. \& Everitt, C.W. \& Bye, M. : 1989. 
{\it NASA, Relativistic Gravitational Experiments in Space}.
The Stanford equivalence principle program.
\bibitem{70}
Nieto, M.M. \& Turyshev, S.G. : 2003. {\it arXiv:gr-qc/0308017}. 
Finding the Origin of the Pioneer Anomaly.
\end{thebibliography}
\end{document}